\documentclass{article}
\usepackage[applemac]{inputenc}
\usepackage{lmodern}
\usepackage{braket}
\usepackage{authblk}
\usepackage{amsfonts}
\usepackage{amsmath}
\usepackage{amssymb}

\title{Bohmian Classical Limit in Bounded Regions}
\author{Davide Romano}
\affil{Department of Philosophy, University of Lausanne UNIL\\CH-1015 Lausanne\\(davide.romano@unil.ch)}
\date{}

\begin{document}
\maketitle

\begin{abstract}
Bohmian mechanics is a realistic interpretation of quantum theory. It shares the same ontology of classical mechanics: particles following continuous trajectories in space through time. For this ontological continuity, it seems to be a good candidate for recovering the classical limit of quantum theory. Indeed, in a Bohmian framework, the issue of the classical limit reduces to showing how classical trajectories can emerge from Bohmian ones, under specific classicality assumptions.
\\In this paper, we shall focus on a technical problem that arises from the dynamics of a Bohmian system in bounded regions; and we suggest that a possible solution is supplied by the action of environmental decoherence. However, we shall show that, in order to implement decoherence in a Bohmian framework, a stronger condition is required (\emph{disjointness of supports}) rather than the usual one (\emph{orthogonality of states}).
\end {abstract}

\section{Bohmian mechanics and classical limit}

Despite the great success of quantum mechanics, a rigorous and general account of the classical limit has not yet been reached. This means we do not have a clear explanation of the transition from the quantum regime, which describes the short-scale world, to the classical regime, which describes our familiar macroscopic world. \\We know that quantum mechanics is a fundamental theory: it applies at every scale.\footnote{Indeed, it is possible to have macroscopic quantum effects, like superconductivity.}The goal of the classical limit, therefore, is to derive classical mechanics from quantum mechanics, under specific classicality conditions.\footnote{The classicality conditions are the physical conditions that allow for the emergence of a classical regime. For example, in decoherence theory, the classicality condition is the (ubiquitous) entanglement of quantum systems.} \\The problem here is not only mathematical, but also conceptual: in standard quantum mechanics (SQM), the physical state of an N-particle system is described by a state vector, an element of an abstract Hilbert space. \footnote{If the state vector is expressed in the position basis, then we have the wave function of the system, which is defined over the 3N-dimensional configuration space of the system.} Moreover, in SQM the state vector has just such a statistical character: for a 1-particle system, the absolute square of the wave function has the meaning of a probability density for finding the particle in a definite region if we perform a position measurement on the system. Within this framework, even if we succeeded in deriving the classical equations of motion for a quantum system, should we regard this result as a true classical limit? Probably not. Classical mechanics describes the motion of particles in space, i.e., it describes real paths for the systems (trajectories) and not just 'probability amplitude' paths. How can we derive the former dynamical structure (and ontology) starting from the latter one?\footnote{See, for example, Holland (1993, sect. 6.1) on the conceptual difference between a quantum 'trajectory' and a classical one.}
\\One option is to consider Bohmian mechanics (BM) as the correct interpretation of quantum theory. In BM, a quantum system is described by a wave function together with a configuration of particles, each following a continuous trajectory in 3D physical space. Within this framework, both quantum systems and classical systems are composed by matter particles that follow real paths in 3D space.\footnote{Of course, in BM there is something more: the wave function. Whether the wave function in BM is a real physical entity (i.e., a new physical field) or a nomological entity that only describes how the particles move (the analogy is with the Hamiltonian in classical mechanics) is currently a subject of philosophical debate. Supporters of the first view are, e.g., Holland (1993) and Valentini (1992); supporters of the second view are, e.g., D\"{u}rr, Goldstein \& Zangh\`{\i} (2013), Goldstein \& Zangh\`{\i} (2012) and Esfeld \textit{et alii} (2014).} So that the entire issue of the classical limit reduces to the question: \emph{under which conditions do the Bohmian trajectories become Newtonian}?
\\However, one could object that classical mechanics is just a high-level effective theory and that the very concept of a 'particle' does not belong to the ontology of the fundamental physical world. In quantum field theory (QFT), for example, the concept of a particle might play no role.\footnote{See, e.g., Malament (1996).} If we cannot introduce a particle ontology at the level of QFT, then we might not see the necessity of introducing it at the non relativistic quantum level either: a characterization of the theory in terms of the wave function could also be enough for QM. Under this view, the classical limit is obtained by the description of a narrow wave packet following a classical trajectory.\footnote{We note that, within the SQM framework, this approach seems to miss the conceptual point of the classical limit problem. In SQM, the wave function is not a real entity, but mainly a mathematical tool for extracting probabilities of the measurement outcomes. Therefore, a narrow wave packet that follows a classical trajectory simply means that whenever we perform a position measurement on the system, we will obtain a result that is compatible with a classical trajectory. Nonetheless, we cannot extract the picture of a real entity following a classical trajectory from that. In other words, what is problematic is not considering a narrow wave function as a particle, but the statistical interpretation of the wave function as opposed to a real ontological entity (particle) following a trajectory in space.} This is the standard approach we usually find in SQM textbooks\footnote{See, e.g., Merzebacher (1970, Ch. 4), Shankar (1994, Ch. 6), Sakurai (1994, Ch. 2). In particular, Shankar sheds some light on specific limitations of the theorem.}, known as \emph{Ehrenfest's theorem}.  
\\However, it is worth noting that some specific QFT models with a particle ontology have been proposed\footnote{See D\"{u}rr \textit{et alii}(2004)}, so that the philosophical inquiry about the fundamental ontology of the physical world is still open.
\\Nevertheless, Ehrenfest's theorem alone cannot provide a proper solution for the quantum to classical transition. First, the wave function of an isolated quantum system generally spreads out in a very short time. Moreover, Ballantine shows that Ehrenfest's theorem is neither necessary nor sufficient for obtaining a classical dynamical regime for quantum systems\footnote{See Ballantine (1994), (1996), (1998, sect. 14.1).}. 
\\The most convincing approach for the analysis of the quantum to classical transition is actually decoherence theory. So, in order to find out how Newtonian trajectories can emerge from Bohmian ones, it seems reasonable to check whether and how decoherence theory fits into the Bohmian framework.
\\The aim of the paper is to focus on a technical problem, which arises in the context of BM in the attempt to derive classical trajectories for a pure state system in bounded regions. The problem follows from the fact that two (or more) Bohmian trajectories of a system cannot cross in the configuration space of the system. So, even if we assume that a macroscopic body, satisfying some specific-classicality conditions (big mass, short wavelength, etc.), starts following at the initial time a classical trajectory, its motion will become highly non-classical if, at a later time, different branches of the wave function of the body cross each other in configuration space.
\\We argue that a possible solution is offered by the action of environmental decoherence on the system.\footnote{This solution was originally proposed by Allori \textit{et alii} (2002).} A relevant point will be clear from the analysis: in order to implement decoherence in the framework of BM, a stronger condition is required (\emph{disjointness of supports}) than the usual one (\emph{orthogonality of states}) for systems describing the environmental particles that scatter off the (macroscopic) Bohmian system.
\\In section 2, we will describe the measurement process in BM, focusing on the emergence of the \emph{effective wave function}. In section 3, we will present the problem mentioned above, which arises (mainly) in bounded regions. In section 4.1, we will introduce decoherence theory as the crucial ingredient for the quantum to classical transition in every physically realistic situation. In section 4.2, we will show how a simple model of environmental decoherence can solve the problem, thus leading to the emergence of classical trajectories in bounded regions.

\section{Bohmian mechanics}
\subsection{A short introduction to Bohmian mechanics}

Bohmian mechanics is a quantum theory in which the complete physical state of an N-particle system is described by the pair $(Q,\Psi)$, where $Q=(q_1, q_2, \ldots, q_N)$ is the configuration of N particles, each particle $q_k (k=1, 2, \ldots, N)$ living in 3D physical space,\footnote{Thus, the configuration $Q$ is defined over the 3N-D configuration space of the system.} and $\Psi=\Psi(Q,t)$ is the wave function of the system, which is defined over the 3N-D configuration space of the system. For a non-relativistic spinless N-particle system, the dynamical evolution of the Bohmian system is given by the Schr\"{o}dinger equation:
$$ i\hbar\frac{\partial{\Psi(Q,t)}}{\partial{t}}=-\sum_{k=1}^{N}\frac{\hbar^2}{2m_k}\nabla_k^2{\Psi(Q,t)}+V\Psi(Q,t)$$ 
which describes the time evolution of the wave function, and the guiding equation:
\begin{displaymath}
 \frac{dq_k}{dt}=\frac{\hbar}{m_k}\mbox{Im}\frac{\nabla_k{\Psi(Q,t)}}{\Psi(Q,t)}  \mbox{ ;  } \mbox{ with } k=1, 2, \ldots, N
 \end{displaymath}
which describes the time evolution of each particle position of the total configuration. From the guiding equation, we note the non-local dynamics of the Bohmian particles: the velocity of a single particle $(q_k)$ will depend on the position of all the other particles of the total configuration $(Q=(q_1, q_2, \ldots, q_N))$. For obtaining a successful scheme of the quantum to classical transition, we need to explain not only the emergence of classical trajectories but also the passage from a quantum (holistic) non-local dynamics to a classical (separable) non-local dynamics.\footnote{In classical mechanics, the potentials that affect the particle motion decay quadratically with the distance, so that we can effectively describe the motion of one particle as autonomous and independent from the motion of a very distant particle (under specific conditions, of course). In BM, instead, the influence of the ``quantum potential'' on the particle motion does not decay with the distance, so that all the particles belonging to the configuration of a system are holistically related, even if they are located very far away from each other. See, e.g., Bohm (1987, sect. 3) for a clear explanation about the difference between quantum (Bohmian) and classical non-locality.}
\\Bohmian mechanics introduces quantum probabilities as a measure of subjective ignorance on the initial conditions of a system (\emph{epistemic probabilities}): given a system with wave function $\psi$, our maximum knowledge about the actual initial positions of the particles is represented by a statistical distribution of possible configurations, i.e., a \emph{classical ensemble}, according to the absolute square of the wave function: $$\rho(Q)=|\psi(Q)|^{2}$$ This is a postulate in BM and it is known as \emph{quantum equilibrium hypothesis}.\footnote{The justification of the quantum equilibrium hypothesis is a subtle issue. Two main approaches have been proposed: the typicality approach by D\"{u}rr, Goldstein \& Zangh\`{i} (1992) and the dynamical relaxation approach by Valentini (1991).} Moreover, from the Schr\"{o}dinger equation, it follows that $\rho$ has the property of equivariance: $$ \mbox{if } \rho(Q,0)=|\psi(Q,0)|^{2}\mbox{, then } \rho(Q,t)=|\psi(Q,t)|^{2} \mbox{ ; }\forall t>0 $$  
\emph{Quantum equilibrium} and \emph{equivariance} imply that BM provides the same empirical predictions of SQM, once it is assumed that the result of a measurement is always encoded in a definite position of a pointer\footnote{We call \emph{pointer} every measurement apparatus that shows a definite outcome after the physical interaction with a quantum system.} and that different positions of a pointer are always represented by (approximately) non-overlapping supports in configurations space.\footnote{We will analyze this condition in more detail in the next section.}

\subsection{Measurement processes in Bohmian mechanics}

In this section we analyze a typical measurement process in BM, showing, in particular, how an \emph{effective wave function} in a Bohmian system emerges. Then we will show that the condition of disjoint supports for different positions of a pointer is essential for obtaining a clear and definite measurement result. 
\\ Let's consider a system $\Psi(x)$, with actual configuration $X$, interacting with an apparatus $\Phi(y)$, with actual configuration $Y$.\footnote{ Bohmian systems are always composed of a wave function and real particles, each of them having a definite position in space. We call \emph{actual configuration} the configuration of particles described by their definite positions in space, and mathematically expressed by $Q=(q_1, q_2, ..., q_N)$. } We suppose that the degrees of freedom of the system and the apparatus are respectively $m$ and $n$, such that the support of $\Psi(x)$ is defined over the $m$-dimensional configuration space of the system and the support of $\Phi(y)$ over the $n$-dimensional one of the apparatus\footnote{A \emph{support} of a function is the region of its domain in which it is not zero valued.}. We suppose that the initial state of the system is a superposition of two wave functions: $$\Psi(x)=\alpha\psi_1(x)+\beta\psi_2(x)$$ with normalization $|\alpha|^{2}+|\beta|^{2}=1$. 
\\At the initial time $t=0$, the system and the apparatus have not yet interacted, so the wave function of the total system (system + apparatus) is factorized: $$\Psi(x,0)\Phi(y,0)=(\alpha\psi_1(x,0)+\beta\psi_2(x,0))\Phi(y,0)$$
\\During the time interval $\Delta t=(0,T)$, the system and the apparatus will evolve according to the Schr\"{o}dinger equation: in a typical measurement interaction, thanks to some coupling term between the two, they will become entangled: $$\Psi(x,0)\Phi(y,0) \longrightarrow \alpha\Psi_1(x,T)\Phi_1(y,T)+\beta\Psi_2(x,T)\Phi_2(y,T)$$ 
This is the usual formulation of the measurement problem: the physical state of the total system, after the measurement, represents a coherent superposition of two macroscopically distinct pointer states. In BM, there is a further ingredient that permits the (dis)solution of the problem: besides the wave function, every Bohmian system is composed of an actual configuration of particles. So, after the measurement interaction, the macroscopic pointer will show a unique and definite result, embodied by the configuration of particles that compose the pointer. In other words, it is the evolution of the particles that finally determines which one of the possible pointer states (described by the evolution of the wave function) will be realized during the measurement process.
\\We suppose, for example, that $\phi_1$ is the wave function corresponding to the physical state of the pointer ``pointing to the left'' and $\phi_2$ that of the pointer ``pointing to the right'': at the time $t=T$, if $Y\in\mbox{\emph{supp}}(\phi_1)$, then the pointer points to the left, if $ Y\in\mbox{\emph{supp}}(\phi_2)$, then it points to the right. Since the two supports are (macroscopically) disjoint,\footnote{It is worth noting that the concept of a perfect \emph{disjointness of supports} is an idealization: the support of a wave function is typically unbounded in configuration space. As a first approximation, we can say that two different supports are disjoint if they have negligible overlap in configuration space. More precisely, we will say that the supports of two different wave functions are (macroscopically) disjoint when their overlap is extremely small in the square norm over any (macroscopic) region. } i.e.,$\mbox{\emph{ supp}}(\phi_1) \cap \mbox{\emph{supp}}(\phi_2)\cong\emptyset$ , then the final result is unique and the superposition disappears.\footnote{The idea is that, since different macroscopic states of the pointer occupy different regions in 3D physical space, the wave functions describing these states will have disjoint supports in the 3N-D configuration space of the pointer.}
\\Suppose, for example, that after the interaction between the system and the apparatus, $Y\in\mbox{\emph{supp}}(\phi_1)$: in this case, the actual configuration of the particles that compose the apparatus will be arranged in space in such a way as to form a physical pointer pointing to the left. Moreover, because of the entanglement\footnote{During the interaction, the dynamics of the particles of the system is strongly related to that of the particles of the apparatus, so that if $Y\in\mbox{\emph{supp}}(\phi_{1(2)}), \mbox{ then } X\in\mbox{\emph{supp}}(\psi_{1(2)})$.} between the system and the apparatus during the interaction, the actual configuration of the particles that compose the system will be in the support of $\psi_1$, that is, $X\in\mbox{\emph{supp}}(\psi_1)$. In this case, we will say that $\psi_1$ is the \emph{effective wave function} (EWF) of the system, i.e., the branch of the total superposition that contains and guides the particles of the system after the interaction, whereas $\psi_2$ is the \emph{empty wave function}, which can be FAPP\footnote{For All Practical Purposes (acronym introduced by John Bell)} ignored after the interaction.  
\\Assuming the quantum equilibrium hypothesis and the condition of disjoint supports any two different pointer states, it is easy to show that the probability distribution of the measurement outcomes is given according to Born's rule. For example, in the case discussed above, we see that the probability of getting the eigenvalue associated with the eigenfunction $\phi_1$ in a measurement is:\footnote{We follow here the derivation presented in D\"{u}rr \& Teufel (2009, sect. 9.1).} 
\begin{displaymath}
\begin{split}
P(Y(t=T)\in supp(\phi_1))=\\ =\int_{\mathbb{R}^m}d^mx \int_{supp(\phi_1)}d^ny |\alpha\psi_1(x,T)\phi_1(y,T)+\beta\Psi_2(x,T)\phi_2(y,T)|^{2} =\\ =\int_{\mathbb{R}^m}d^mx \int_{supp(\phi_1)}d^ny|\alpha\psi_1(x,T)\phi_1(y,T)|^{2} +\\ +\int_{\mathbb{R}^m}d^mx \int_{supp(\phi_1)}d^ny |\beta\psi_2(x,T)\phi_2(y,T)|^{2}+ \\+ 2\Re  \int_{\mathbb{R}^m}d^mx \int_{supp(\phi_1)}d^ny  \mbox{ }\alpha\beta^*\psi_1(x,T)\psi_2^*(x,T)\phi_1(y,T)\phi_2^*(y,T)\cong \\ \cong|\alpha|^{2} 
\end{split}
\end{displaymath}
 \\which is in agreement with the Born's rule.\footnote{A specular derivation can be made for the other possible outcome of the measurement: in this case we need to integrate in the support of $\phi_2$ and the final probability will be $|\beta|^2.$.}
\\In the derivation we have used the quantum equilibrium hypothesis for the first equation and $$\int_{supp(\phi_1)}d^ny|\phi_2(y,T)|^{2}\cong 0 $$ $$\int_{supp(\phi_1)}d^ny\mbox{ }\phi_1(y,T)\phi_2(y,T)\cong0$$ \\because $supp(\phi_1)\bigcap supp(\phi_2) \cong \emptyset$.
\\The emergence of the effective wave function of the system $\psi_1(x,T)$ represents a first step in the transition from a holistic regime to a local one:\footnote{By \emph{holistic} we mean the quantum (Bohmian) non-locality, with \emph{local} the classical non-locality. This terminology was introduced by Esfeld \textit{et alii}(2014)(forthcoming).} after the measurement, the initial superposition of the total system effectively collapses\footnote{In BM, there is never a real collapse of the wave function.} in just one of the possible branches, which is described by a factorized state between an eigenfunction of the system and one of the apparatus, e.g., $\psi_1(x,T)\phi_1(y,T)$. Hence, the dynamics of the system is now decoupled from that of the apparatus: the further evolution of the particles of the system will be autonomous and independent from that of the particles of the apparatus (because now they belong to distinct and factorized wave functions). Moreover, interference with the empty wave function will be practically impossible, given the condition of disjoint supports for the wave functions of different pointer states.
\\We might say that the EWF describes a \emph{local} dynamics of the system, since the particle evolution of the sub-system described by $\psi_1$ does not depend on the position of the particles of any external system. Whenever an EWF emerges, the holistic Bohmian non-locality seems, at least temporarily, to be turned off. 
\\A simple example can help us visualize the situation. Let's consider a typical EPR set up: generally, changing some potentials on one wing of the system, say at point A, will influence the trajectory of the particle on the other wing, say at point B.\footnote{We suppose that points A and B are space-like separated.} Nevertheless, if, as a consequence of a measurement, an effective wave function emerges (e.g., at point B), then the trajectory of the particle on the B-side can be influenced only by potentials on its side (i.e., potentials that are connected with B by time-like intervals).
\\Of course, this is only a first step towards the classical world. The other important step is to show how classical trajectories can emerge starting from Bohmian ones.\footnote{In the following, we will not face the problem of the emergence of classical trajectories in BM. The interested reader should see, e.g., Rosaler (2014), for a \emph{decoherent histories} approach to the Bohmian classical limit; Appleby (1999) and Sanz, Borondo (2004), for an analysis of specific models where Bohmian trajectories, implemented in a regime of full decoherence, become classical.} In the next section, we will discuss a technical problem arising for the Bohmian classical limit in bounded regions and we will see how decoherence can solve the problem. In section 4, we will briefly introduce decoherence and, finally, we will clarify the mathematical conditions for implementing it in the framework of BM. 

\section{Bohmian classical limit in bounded regions}
In this section, we focus on a problem that arises from the dynamics of a Bohmian system in bounded regions.\footnote{For the sake of clarity, the problem can also arise in unbounded regions: indeed, it is a consequence of a simple mathematical fact, so it is fundamentally independent from the nature (bounded or unbounded) of the space in which the system moves. Nevertheless, since it is more likely to happen in bounded regions than unbounded ones, then it seems more natural to set the problem in a bounded region.} The problem was originally discussed in Allori \textit{et alii} (2002, sect. 8). However, for the sake of completeness, we will briefly restate it here.
\\We consider an infinite potential well of size $L$ in one dimension and a 1-particle Bohmian system in the center of the well. We suppose that the wave function of the system is a linear superposition of two wave packets with opposite momenta. In the classical limit model, the position $x$ of the system will be the center of mass of a macroscopic body whose classical motion we are searching for.   
\\At the initial time $t=0$, we suppose that the two packets begin to move classically in opposite directions.\footnote{We suppose, to start with, classical trajectories for each branch of the wave function, which is equivalent to assume a classical limit in unbounded regions. In this regard, some partially successful results have already been achieved (I briefly indicate the main approach adopted by the authors for each reference): Allori \textit{et alii} (2002): quantum potential \emph{plus} Ehrenfest's theorem; Holland (1993, Ch. 6): quantum potential; Bowman (2005): mixed states \emph{plus} narrow wave packets \emph{plus} decoherence; Sanz \& Borondo (2004) and Appleby (1999): decoherence; Rosaler (2014): decoherent histories.} At time $t_R$, they (approximately)\footnote{The velocity field in BM is never bi-valued, so the particle arrives very close to the well, but without touching it} reach the walls and, for $t>t_R$, they start to converge towards the center. At time $t_c=2t_R$ (\emph{first caustic time}), the two wave packets will cross each other in the middle of the well, but, since the Bohmian trajectories of a system cannot cross\footnote{Bohmian trajectories cannot cross in configuration space because the guiding equation is a first-order equation, so to each position $x$ corresponds a unique velocity vector $v$.} in the configuration space of the system,\footnote{For a 1-particle system, the configuration space of the system corresponds to the 3D physical space.} the two converging trajectories will not cross each other: the trajectory coming from the right-hand side will start to come back to that side after time $t_c$. In a perfectly symmetric way, the same will happen for the trajectory coming from the left-hand side of the well. So, for example, if the particle is contained, at the beginning, in the wave packet that goes to the right, then it will move in future only within the right-hand side of the well. And this is clearly not classical behavior.\footnote{Note that this situation is completely different from the case of the ``surrealistic trajectories'' in BM. In the latter, it is after all not so problematic to have odd trajectories, if they finally match with the empirical predictions of QM. In this case, instead, we want to recover the classical dynamics of a macroscopic body, so the empirical predictions to match with are the trajectories of classical mechanics. Thus, no non-classical trajectory of the system can match with the empirical result we expect from a classical limit model.} 
\\Nevertheless, Allori \textit{et alii} (2002) claim that, in a realistic model, we also need to take into account interaction with the environment, and then the problem should vanish. Indeed, an external particle (a neutrino, a photon, an air molecule,\ldots), interacting with the (macroscopic) system before the caustic time $t_c$, will ``measure'' the actual position of the center of mass of the system, thus eliminating the superposition between the two wave packets of the system. In other words, the interaction between the external particle and the system acts like a position measurement on the system, performed by the ``environment''. Consequently, the environmental interaction will select only one of the two wave packets of the system, which becomes the \emph{effective wave function} of the system.
\\Here the original passage:
\begin{quote}
These interactions --even for very small interaction energy-- should produce \emph{entanglement} between the center of mass $x$ of the system and the other degrees of freedom $y$, so that their effective role is that of ``measuring'' the position $X$ and suppressing superpositions of spatially separated wave functions. (Taking these interactions into account is what people nowadays call decoherence [...]). Referring to the above example, the effect of the environment should be to select [...] one of the two packets on a time scale much shorter than the first caustic time $t_c$. (Allori \textit{et alii}, 2002, sect. 8, p. 12)
\end{quote}
The solution proposed by Allori \textit{et alii} (2002) raises a subtle conceptual issue. As we saw in section 2.2, an EWF emerges in a Bohmian measurement only if the supports of different pointer states are disjoint in configuration space. When the pointer state is a macroscopic state of a classical apparatus, this condition is generally fulfilled. Nevertheless, in the case of interaction with the environment, the pointer states of the ``apparatus'' are the \emph{environmental states} of the external particle. Therefore, this solution seems to work only if the supports of different environmental states of the external particle, after interaction with the macroscopic system, are disjoint in configuration space. So, the question becomes: \emph{is this condition generally satisfied or not?}\footnote{A related, interesting question is: \emph{what happens if the relative environmental states do not have disjoint supports, but are only (approximately) orthogonal in the Hilbert space of the environment?} At the moment, we do have not a rigorous answer to this question.} Indeed, in order to have \emph{effective decoherence}\footnote{By \emph{effective decoherence}, we mean a decoherence process, within the framework of BM, which is able to produce an effective wave function for the system.} in BM, the condition of disjoint supports for different environmental pointer states has to be satisfied.
\\It is important to note that this is a stronger condition than the usual one required by decoherence in the standard framework, that is, the orthogonality of states.
\\In the next section, we will analyze a simple but realistic model of decoherence, namely environmental decoherence induced by scattering. The analysis will clarify the difference between the standard condition and the Bohmian one required for decoherence.

\section{Decoherence approach to the Bohmian classical limit}

\subsection{A short introduction to decoherence}
Decoherence is the local suppression of the phase relations between different states of a quantum system, produced by the entanglement between the system and its environment,\footnote{In general, the environment can be thought either as external or internal degrees of freedom of a (macroscopic) system.} where the latter is also described as a quantum system.
\\We consider a pure state system $\ket{\psi}=\alpha\ket{\psi_1}+\beta\ket{\psi_2} $ and a pure state environmental system $\ket{\xi}$: as long as they do not interact, they remain physically independent and the total wave function is factorized: $$\ket{\Psi_0}=\ket{\psi}\ket{\xi}=(\alpha\ket{\psi_1}+\beta\ket{\psi_2})\ket{\xi}$$ The density operator of the total system can be also factorized into the density operator of the system and that of the environment: $$ \hat{\rho}^{\Psi_0}=\ket{\Psi_0}\bra{\Psi_0}=\ket{\psi}\ket{\xi}\bra{\xi}\bra{\psi}=\hat{\rho}^{\psi}\otimes\hat{\rho}^{\xi}$$ When the system interacts with the environment, the two systems become entangled and they form a new pure state system:
$$\ket{\Psi}=\alpha\ket{\psi_1}\ket{\xi_1}+\beta\ket{\psi_2}\ket{\xi_2}$$ In a realistic physical model, the system will interact (and, then, become entangled) with many environmental states $\ket{\xi_i}$\footnote{A good approximation for 'many' is the Avogadro number $N_A=6,022X10^{23}$.} in a very short time. Tracing out the degrees of freedom of the environment, we obtain the reduced density operator of the system. Under the assumption of the (approximate) orthogonality of the environmental states, which is essentially the standard condition for decoherence, the reduced density operator formally appears as (approximately) describing a mixture of states: $$\hat{\rho}^{\psi}_{red}=Tr_{\xi_{i}}\ket{\Psi}\bra{\Psi}\cong|\alpha|^{2}\ket{\psi_1}\bra{\psi_1}+|\beta|^{2}\ket{\psi_2}\bra{\psi_2} \mbox{ if    } \braket{\xi_i|\xi_j}\cong\delta_{ij}$$
Nevertheless, it is worth noting that $\hat{\rho}^{\psi}_{red}$ does not represent a proper mixture of states,\footnote{A \emph{proper mixture} of states is an epistemic mixture: the system is in one of the states of the superposition, but we do not know which one of them. An \emph{improper mixture} instead, is a mathematical expression that looks like a proper mixture, yet it describes an ontological superposition of states (see, e.g., Schlosshauer (2007, sect. 2.4)).} but an improper mixture, for three main reasons:
\begin{enumerate}
\item In SQM, the physical state of a system is mathematically represented by the state vector of the system: in this case, the state vector is assigned only to the global entangled state between the system and the environment, and we cannot assign an individual quantum state to a subsystem ($\psi$) of a larger entangled system ($\Psi$).
\item In SQM, the reduced density operator just describes the statistical distribution of the possible outcomes for an observer who locally performs a measurement on the system. So it does not carry information about the physical state of the (sub)system \emph{per se}, but only related to the measurements we can perform on it.
\item Decoherence does not select one particular branch of the superposition. All the different branches remain equally real after the action of decoherence: thus, even if the final state of the system looks like a mixture, it is not a proper mixture that can be interpreted in terms of ignorance about the actual state of the system. We might call it an \emph{improper mixture} (see, e.g., Bacciagaluppi (2011, sect. 2.2)).
\end{enumerate}

\subsection{Environmental decoherence induced by scattering}
Taking decoherence as a realistic background for the classical limit, we first introduce the model of environmental decoherence by scattering,\footnote{The model was originally developed by Joos \& Zeh (1985). Recent accounts of the model can be found in Giulini, Joos \textit{et alii} (2003, Ch. 3) and Schlosshauer (2007, Ch. 3).} and, after this, consider whether the Bohmian condition of disjoint supports could reasonably fit into the model. As for the mathematical presentation of the model, we will mainly follow Schlosshauer (2007, Ch. 3). 
\\We consider a system $S$ that scatters off an external environmental particle, represented by $\xi$. At the initial time $t=0$, $S$ and $\xi$ are uncorrelated: 
$$\hat{\rho}_{S\xi}(0)= \hat{\rho}_{S}(0) \otimes \hat{\rho}_{\xi}(0)$$
Representing with $\ket{x}$ the initial state of the center of mass of the system, with $\ket{\chi_i}$ that of the incoming environmental particle, and with $\hat{S}$ the scattering operator, we can represent the effect of the scattering of a single environmental particle on the system as follows: 
$$\ket{x}\ket{\chi_i} \rightarrow \hat{S}\ket{x}\ket{\chi_i}\equiv\ket{x}\hat{S}_{x}\ket{\chi_i}\equiv\ket{x}\ket{\chi(x)}$$
where $\ket{\chi(x)}$ is the final state of the outgoing environmental particle scattered at $x$ on the system.
\\ From the expression above, we see that if the system is represented by a superposition of different position eigenstates, for example $\ket{x}=\sum_i{a_i\ket{x_i}}$, then the environmental state and the system state will become entangled: the scattering process is a measurement-like interaction, which establishes correlations between the two systems. The environmental states that scatter off the system can be considered as pointer states that encode information about the position $x$ of the system. The scattering process transforms the initial density operator\footnote{In the following, $\hat{\rho}$ and $\rho$ represent, respectively, the density operator and the density matrix of a system. In general, the density matrix is the density operator expressed in a particular basis, usually in the position basis (like in this case).} of the composite system:
$$ \hat{\rho}_{S\xi}(0)=\hat{\rho}_{S}(0)\otimes\hat{\rho}_{\xi}(0)=\int{dx}\int{dx^{\prime}}\rho_{S}(x,x^{\prime},0)\ket{x}\bra{x^{\prime}}\otimes\ket{\chi_i}\bra{\chi_i}$$
into the new density operator:
$$\hat{\rho}_{S\xi}=\int{dx}\int{dx^{\prime}}\rho_{S}(x,x^{\prime},0)\ket{x}\bra{x^{\prime}}\otimes\ket{\chi(x)}\bra{\chi(x^{\prime})}$$
Thus, the reduced density operator of the system after the interaction of a single scattering of an external particle on the system is:
$$\hat{\rho}_{S}=Tr_{\xi}\hat{\rho}_{S\xi}=\int{dx}\int{dx^{\prime}}\rho_S(x,x^{\prime},0)\ket{x}\bra{x^{\prime}}\braket{\chi(x^{\prime})|\chi(x)}$$
Representing the result in the (position basis) density matrix, the evolution of the reduced density matrix of the system under the action of the scattering event can finally be summarized as follows:
$$\rho_{S}(x,x^{\prime},0)\stackrel{scattering}{\longrightarrow}\rho_{S}(x,x^{\prime},0)\braket{\chi(x^{\prime})|\chi(x)}$$
This is an important result: in the SQM model of decoherence induced by scattering, the condition for the local suppression of the spatial coherence of the system is given by the orthogonality of the relative environmental states that scattered off the system:
$$\mbox{\emph{Standard condition for decoherence}:} \mbox{     } \braket{\chi(x^{\prime})|\chi(x)}\cong0$$
In a Bohmian model, this condition is not sufficient for effective decoherence. Indeed, during the scattering process, the environmental state (the external particle) becomes entangled with the system (a macroscopic body, in the classical limit), thus acting like a pointer that measures the position of the center of mass of the system. Nevertheless, as we saw in section 2.2, a good measurement interaction\footnote{That is, a measurement providing a definite outcome.} can be realized in BM only if the wave functions of different states of the pointer have disjoint supports in configuration space. Therefore, to obtain a local suppression of the spatial coherence of the system, BM requires the supports of relative environmental states to be disjoint in configuration space. If $\ket{y}$ indicates a generic position eigenstate of the scattered environmental particle, and $\mathcal{Q}_{\xi}$ the configuration space of the environment, then the Bohmian condition of having effective decoherence induced by scattering is:\footnote{This result is not new: see, e.g., Rosaler (2014, sect. 5, eq. 20) and references therein. What we are aiming to clarify here is the strong connection between this result and the measurement process in BM as well as its conceptual consequences in the context of the classical limit in BM. Moreover, while Rosaler (2014, sect. 5) assumes that the Bohmian condition for decoherence is always satisfied (Rosaler's justification mainly relies on the high-dimensionality of $\mathcal{Q}_{\xi}$), we actually don't see any compelling reason for assuming the condition is satisfied for a typical model of environmental decoherence (e.g., in the short-wavelength limit, even a few external particles suffice to produce decoherence, so the high-dimensionality argument of $\mathcal{Q}_{\xi}$ does not hold in this case). We think, instead, that this issue might deserve further analysis, even with the help of some quantitative results.}
$$\mbox{\emph{Bohmian condition for (effective) decoherence}:} \mbox{ } \braket{\chi(x^{\prime})|y}\braket{y|\chi(x)}\cong0 \mbox{ ; } \forall{y}\in{\mathcal{Q}_{\xi}}$$
or, in terms of the wave function of the scattered environmental particle:
$$supp(\psi_{\chi(x)}(y)) \cap supp(\psi_{\chi(x^{\prime})}(y))\cong\emptyset \mbox{ ; with } supp(\psi_{\chi}(y))\in \mathcal{Q}_{\xi}$$
So, the following question arises: \emph{is the condition of disjoint supports verified in a typical realistic model of environmental decoherence by scattering}?
\\In the case of a ``classic'' quantum measurement process,\footnote{That is, when the pointer states correspond to physical states of a classical apparatus.} we have at least two main reasons to believe that the condition of disjoint supports is fulfilled: 
\begin{enumerate}
\item A classical apparatus is made of an extremely high number of (Bohmian) particles, thereby the configuration space of the apparatus is very high-dimensional (proportional to $10^{23}$D). This makes the probability of a significant overlap between the supports of two different macroscopic pointer states very small.\\(\emph{high dimensional configuration space})
\item The wave function of a macroscopic system, like a classical apparatus, is usually very narrow. Moreover, since different macroscopic pointer states occupy different regions in 3D physical space, the wave functions representing these states will be reasonably defined over regions with disjoint supports in configuration space.\\(\emph{narrow wave function})
\end{enumerate}
Nevertheless, the situation changes dramatically when the apparatus is not a macroscopic object, but a microscopic environmental particle, the latter being either a photon, an electron, a neutrino, etc. Indeed, the assumptions mentioned above simply do not apply when the pointer state is a microscopic system:
\begin{enumerate}
\item The wave function of a microscopic system is generally not very narrow, and, moreover, it usually spreads out in configuration space in a very short time.\\(\emph{wave function spreads out})
\item In some limiting cases, we can send only a few particles that scatter off the system to produce decoherence effects (this is generally true, for example, in the short-wavelength limit.\footnote{See, e.g., Schlosshauer (2007, sect. 3.3.1) and Joos \textit{et alii}(2003, sect. 3.2.1.1).}) In this case, the configuration space of the environment $\mathcal{Q}_{\xi}$ is not very high-dimensional.\\(\emph{low-dimensional configuration space})
\end{enumerate}
Since the traditional arguments\footnote{See, e.g., D\"{u}rr \& Teufel (2009, sect. 9.1). It is worth noting that in section 9.2 these authors generalize the quantum measurement process by including the case in which the pointer is a microscopic system. They affirm that it is precisely thanks to decoherence processes that an effective wave function is produced ``more or less all the time, and more or less everywhere''. We agree with them in considering entanglement and decoherence essential for the production of effective wave functions and for the emergence of a (classical non-) local world. Nevertheless, their arguments for the validity of the condition of disjoint supports in the case when the pointer is a microscopic system are pretty qualitative, so they cannot be viewed as a definitive answer to this problem.} for the validity of the condition of disjoint supports do not apply when the measurement apparatus is a microscopic quantum system (such as an environmental particle), and \textit{prima facie} we do not have any strong argument for considering the condition satisfied, the question remains open and worthy of future work.
\\Let us now offer some final (and more speculative) remarks on the conceptual consequences of the analysis of the conditions for Bohmian decoherence. We note that if the condition of disjoint supports is generally satisfied in a typical model of environmental decoherence, then decoherence fits very well in the framework of BM. Yet BM could account for the selection of just one trajectory within the branching linear structure produced by the Schr\"{o}dinger evolution of open quantum systems, without the need for a real collapse of the wave function at some stage of the process (SQM) or the introduction of many simultaneous non-detectable existing worlds (Everett, MWI). 
\\On the other hand, if the condition of disjoint supports is not generally satisfied in those models, then it may be possible to find some regime in which BM gives different empirical predictions from SQM. Let's consider, for example, a decoherence model in which the condition of orthogonality of states is satisfied, whereas the condition of disjoint supports is not. Under this model, SQM and BM will predict different phenomena: according to SQM, we will obtain decoherence effects; according to BM, we will not. Suppose that we were able to realize an experimental set up that physically implemented this model. Performing the experiment, we will hypothetically be able to distinguish whether SQM or BM is true, since the two theories provide different empirical predictions under the same model. Of course, things might be not so simple, for many reasons. First, we should write a mathematical model in which the condition of orthogonality of states and that of disjoint supports come apart. Second, the model should be practically implementable into a real physical set up. In any case, what we find interesting is that, if the condition of disjoint supports is really necessary for implementing decoherence in BM, then the possibility is open for finding (at least hypothetically) some physical regimes where the Bohmian empirical predictions are different from the SQM ones. 

\section{Conclusions}
Decoherence theory is the standard framework for showing how classical trajectories and classical states can emerge from the quantum world, and it is a crucial ingredient in BM in order to recover the emergence of classical trajectories in bounded regions.\\
We showed that, in order to implement an \emph{effective decoherence} in BM, i.e., a physical mechanism that gives rise to an effective wave function for a Bohmian system through interaction with the environment, a condition stronger than the standard orthogonality of states is required: the supports of relative environmental states have to be disjoint in the configuration space of the environment.
\\Thus, a relevant open issue for recovering the classical limit in BM is to verify whether this condition is satisfied for typical realistic models of environmental decoherence.

\section*{Acknowledgments}
I would like to express special thanks to Guido Bacciagaluppi and Antonio Vassallo for their helpful comments on earlier drafts of the paper, and for a continuous exchange of ideas about this topic. Their suggestions very much improved my original work.
\\I am also indebted to the philosophy of physics group in Lausanne: Michael Esfeld, Vincent Lam, Matthias Egg, Andrea Oldofredi, and Mario Hubert, for helpful comments and discussions.
\\This work was supported by the Swiss National Science Foundation through the research project ``\emph{The metaphysics of physics: natural philosophy}''.

\end{document}